 \documentclass[smallabstract,smallcaptions]{dccpaper}

\usepackage{epsfig}
\usepackage{citesort}
\usepackage{amsmath}
\usepackage{amssymb}
\usepackage{color}
\usepackage{url}
\usepackage{cite}
\usepackage{adjustbox}
\usepackage{algorithm}
\usepackage{algpseudocode}
\usepackage{booktabs}
\usepackage{multirow}
\newlength{\figurewidth}
\newlength{\smallfigurewidth}
\newcommand{\Lagr}{\mathcal{L}}

\newcommand\blfootnote[1]{
    \begingroup
    \renewcommand\thefootnote{}\footnote{#1}
    \addtocounter{footnote}{-1}
    \endgroup
}

\begin{document}

\title
{\large
\textbf{A Neural Enhancement Post-Processor with a Dynamic AV1 Encoder Configuration Strategy for CLIC 2024}
}

\author{%
Darren Ramsook$^{\ast}$ and Anil Kokaram$^{\dag}$\\[0.5em]
{\small\begin{minipage}{\linewidth}\begin{center}
\begin{tabular}{ccc}
Sigmedia Group,\\
Department of Electronic \& Electrical Engineering,\\
Trinity College Dublin,\\
Dublin, Ireland.\\
$^{\ast}$\url{ramsookd@tcd.ie}, $^{\dag}$\url{anil.kokaram@tcd.ie} 
\end{tabular}
\end{center}\end{minipage}}
}

\maketitle
\thispagestyle{empty}

\begin{abstract}
At practical streaming bitrates, traditional video compression pipelines frequently lead to visible artifacts that degrade perceptual quality. This submission couples the effectiveness of a neural post-processor with a different dynamic optimsation strategy for achieving an improved bitrate/quality compromise. The neural post-processor is refined via adversarial training and employs perceptual loss functions. By optimising the post-processor and encoder directly our method demonstrates significant improvement in video fidelity. The neural post-processor achieves substantial VMAF score increases of +6.72 and +1.81 at bitrates of 50 kb/s and 500 kb/s respectively.

\end{abstract}

\Section{Introduction}

There has been an exponential growth in the consumption and distribution of digital video content due to the proliferation of video streaming and teleconferencing platforms~\cite{stocker2023covid}. Video compression has become an essential component of the digital ecosystem. "Lossy" compression compression techniques enable efficient storage, transmission, and delivery. However, at practical bitrates and with increasing picture sizes, it introduces visual artifacts and compromises the overall quality of the compressed video~\cite{9035388}. As a result, there is a pressing need for effective methods to enhance the quality of compressed videos and remove compression artifacts while preserving important details and maintaining the fidelity of the original content.

\blfootnote{© 2023 IEEE.  Personal use of this material is permitted.  Permission from IEEE must be obtained for all other uses, in any current or future media, including reprinting/republishing this material for advertising or promotional purposes, creating new collective works, for resale or redistribution to servers or lists, or reuse of any copyrighted component of this work in other works.}

Generative Adversarial Networks (GANs) have emerged as a powerful tool in image-related tasks, including denoising\cite{tran2020gan, chen2020dn} and super-resolution\cite{wang2018esrgan, Bulat_2018_ECCV}. As such, researchers have naturally turned to GANs for addressing the challenges of compression artifact removal in still images\cite{wang2021real, Galteri_2017_ICCV}. Despite the effectiveness of GANs in still image compression artifact removal, their application to video enhancement is still in its early stages. Existing GAN-based architectures~\cite{zhang2020enhancing, ma2020cvegan, 9413095} for post-processing compressed video enhancement focus solely on processing individual frames in isolation, disregarding the temporal information present in videos. This approach overlooks the inherent temporal dependencies among frames, which play a crucial role in capturing and reproducing the motion patterns and coherent structures in videos. Consequently, the generated videos often exhibit temporal inconsistencies, motion artifacts, and a lack of temporal smoothness. Other neural based approaches for video processes entire sequences at a time~\cite{ho2022video, ho2022imagen}. While this approach allows for robust temporal connections to be made across frames, this approach is limited by the memory of the hardware and uses 3D convolutions which are much more computationally expensive.

As first observed by Katsavounidis et al~\cite{Katsavounidis_2018}, it is possible to select a bitrate/quality operating point (\textit{the encoded representation}) by considering the creation of the bitrate ladder itself as an optimisation task. We present an alternative strategy by using a direct search technique that incorporates the specification of the target bitrate as a parameter as well. There has been significant work that shows the value of optimising a neural pre-processor as part of the pre-processor/encoder pipeline\cite{9487490, Chadha_2021_CVPR, guleryuz2021sandwiched}. In general, post-processors are designed with respect to different encoders but not in conjunction with encoded representations. We therefore develop a scheme that optimises each neural post-processor for every specific representation and associated parameterisation. A key observation here is that in a practical application, constant bitrate (CBR) encoding is employed to generate representations. However in single-pass CBR encoding, the output bitrate rarely achieves the desired target. We explore a method for achieving this bitrate by altering the target bitrate paremeter of an AV1 iteratively in a semi-multipass encoding scheme.

\begin{figure}
    \centering
    \includegraphics[width=\textwidth]{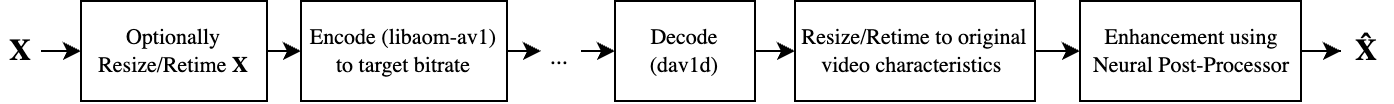}
    \caption{\em Encoding and Decoding Stages of our proposed process. Our resizing/retiming step before encoding is done to ensure the input video meets a specific bitrate. We train multiple neural post-processors which are dependent on the amount of downsampling which is applied to the input video}
    \label{fig:processoverview}
\end{figure}

\begin{figure}
    \centering
    \includegraphics[width=0.7\textwidth]{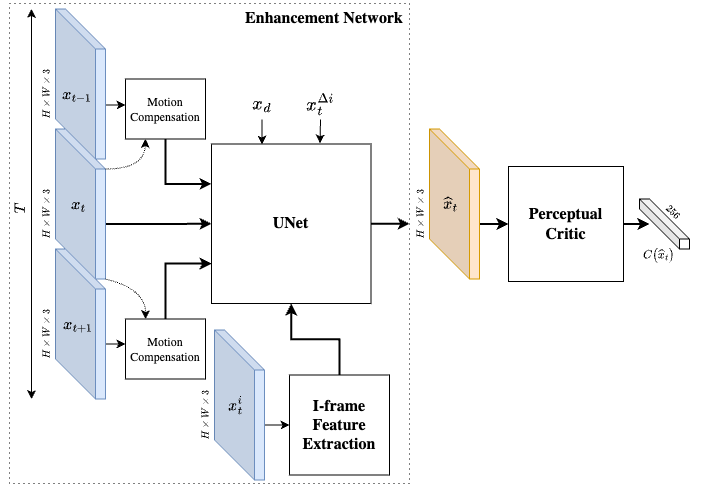}
    \caption{\em Neural Post-processing network. The enhancement network (within the dotted lines) takes three motion compensated frames as input. It also uses the nearest I-frame, a degradation strength, $x_d$, and the distance between the current frame and the nearest I-frame $x_t^{\Delta}i$. The perceptual critic is based of the architecture of~\cite{10222397}.}
    \label{fig:neuraloverview}
\end{figure}

\textbf{Our Contributions:} In this work, we deploy libaom-av1 (version \textbf{3.6.1}), an open-source reference encoder of the AV1 standard\cite{AV1Spec}, as the foundation of our video compression pipeline outlined in Figure~\ref{fig:processoverview}. For decoding, the dav1d (with commit id \textbf{58afe4}) decoder is used followed by our neural post-processor. 
We present four key components.
\begin{enumerate}
    \item A specification of an encoding step as in figure~\ref{fig:processoverview} in which the input content is downsampled spatially and temporally and coupled with a bitrate target parameter to achieve a target bitrate. 
    \item A strategy for achieving a target encoded bitrate by selecting the optimal resolution, frame rate and target bitrate parameter in the actual encoder invocation.
    \item Use of an adversarially trained post-processor incorporating both spatial and temporal frame information as well as perceptual loss criteria. (See Figure~\ref{fig:neuraloverview})
    \item Selecting an optimal post-processor for each different encoder parameterisation.
\end{enumerate} 
\vspace{-1em}

\section{Related Work}

There has been substantial work on post-processor applications for improving compressed still images. The use of perceptual metrics in the loss functions of neural-networks has been proven to have better human subjective quality. In ~\cite{mohammadi2022perceptual}, results indicated that training with either a DFQM or MS-SSIM has the highest perceptual gain. In our training setup, we use the DFQM LPIPS as part of our generator loss function. Using the difference of feature maps from intermediate layers of pre-trained classification networks as a loss has also been shown to give improved results in image tasks~\cite{johnson2016perceptual, 10.1007_978-3-030-05792-3_7}. In~\cite{wang2021real}, the use of this loss has shown to give increased performance in JPEG compression reduction. We include this loss term when training our proposed adversarial setup. 

Previous models for video post-processing for compression artifact removal do not exploit temporal information. The study conducted by~\cite{zhang2020enhancing} introduced a neural-based model for video enhancement, which demonstrated improved Peak Signal-to-Noise Ratio (PSNR) and Video Multimethod Assessment Fusion (VMAF) scores when compared to videos without post-processing. Their approach focused on enhancing the visual quality of videos through a neural network with multiple residual blocks. 

In~\cite{ma2020cvegan}, a post-processing adversarial approach is presented, which incorporates a mixture of multiple objective metrics including SSIM and MSSIM in its loss function. Their approach includes the direct comparison of complete deep features between a degraded-reference pair, similar to~\cite{wang2021real}. This model shows substantial improvement in PSNR and VMAF.

However in~\cite{zhang2020enhancing,ma2020cvegan}, the post-processing networks employed did not utilize temporal information across frames. Instead, it primarily focused on enhancing individual frames independently. While this approach led to enhanced visual quality metrics, the potential benefits of incorporating temporal information and exploiting the correlations between consecutive frames were not fully explored.

\section{Method}

Our compression pipeline is shown in Figure~\ref{fig:processoverview}. The first block relates to resizing the video to be encoded either spatially or temporally. The selection of the amount of downsampling to be used is important to maintain the highest quality without exceeding a given bitrate.

\textbf{Choosing the encoded representation}: We use the libaom-av1 encoder as the foundation for our pipeline. We denote the input video clip as $\mathbf{X}$. Under a CBR invocation of the encoder we specify a target bitrate $r$kb/s and hence generate an output compressed version of $\mathbf{X}$ which when decoded yields $\mathbf{X}^{c}$. The actual encoded bitrate of the compressed stream is $r^c$kb/s. In general $r \neq r^c$ because of various suboptimal choices made in CBR encoding in a practical encoder. However, by altering the requested $r$kb/s we can achieve an output $r^c \approx r$. For example, if $r^c > r$ then we can re-encode with $r_{2} = r  - \delta$ where $r_2$ is the bitrate request on our second invocation of the encoder and $\delta$ is to be estimated. In this work we use the method of bisection to estimate $\delta$ and hence achieve the requested rate $r$ iteratively. We use a maximum of 8 iterations. 


After the maximum iterations are reached it is still possible that we do not achieve the requested bitrate. In that case, $\mathbf{X}$ is downsampled spatially and the target bitrate search process is repeated. We employ up to 3 dyadic downsampling steps (2, 4, 8). 

Under severe rate constraints, where the video is downscaled by a factor of eight and no candidate represenation is realized, the temporal resolution is correspondingly reduced by discarding every other frame, effectively downsampling the frame rate. 

The algorithm for searching across spatial dimensions is presented in Algorithm~\ref{alg:bitrate}. The representation chosen as the encoded format is the version that has the maximum 'PSNR HVS'~\cite{psnrhvs}. Note that to calculate the quality loss across the pipeline we compare the orginal clip with the decoded and upsampled clip. Spatial upsampling is achieved with a 5-tap Lanczos filter. Temporal upsampling where needed is achieved by repeating frames (zero-order hold) to return the clip to its initial frame count. This ensures that the video retains its visual continuity, despite the aggressive bitrate constraints imposed during compression.

\begin{algorithm}
\caption{Spatial Resizing Target Bitrate Search Algorithm}\label{alg:bitrate}

\begin{algorithmic}
\Require $\mathbf{X},\ \alpha,\ r$ \Comment{$\mathbf{X}$: input video, $\alpha$: \# of search steps, $r$: target kb/s}
\Ensure $\beta = 1, r^{c} = \infty$ \Comment{$\beta$: downsample factor, $r^{c}$: output kb/s}
\Ensure $r^{i} = r, \alpha_s = 0$ \Comment{$r^{i}$: codec input kb/s, $\alpha_s$: current search step}
\While{$\beta <= 8$}
    \While {$\alpha_s < \alpha$}
    \State $\mathbf{x} \gets DS(\mathbf{X}, \beta)$ \Comment{$DS(a, b)$: downsample $a$ by a factor of $b$}
    \State $\mathbf{\hat{x}} \gets AV1(\mathbf{x}, r^{i})$ \Comment{$AV1(a, b)$: encode $a$ with $b$kb/s,  $\mathbf{\hat{x}}$: encoded output}

    \State $\zeta \gets rate(\mathbf{\hat{x}})$ \Comment{$\zeta$: bitrate of $\mathbf{\hat{x}}$}

    \State $M \gets metric(\mathbf{X}, \mathbf{\hat{x}})$ \Comment{Save $M$}
    
    \If {$r^{i} = r$}
        \State $adj \gets r^{i}/2$ \Comment{$adj$: Adjustment to be made to $r^{i}$}
    \ElsIf{$r^{i} \neq r$}
        \State $adj \gets abs((r - r^{i})/2$)
    \EndIf
    \If{$\zeta \geq r$} \Comment{Modify $r^{i}$ depending on $\zeta$}
        \State $r^{i} \gets max(1, r^{i} - adj)$
    \ElsIf{$\zeta < r$}
        \State $r^{i} \gets r^{i} + adj$
    \EndIf
    \State $\alpha_s \gets \alpha_s + 1$
\EndWhile
    \State $\beta \gets \beta \times 2$
\EndWhile
\end{algorithmic}
\end{algorithm}

\textbf{Neural Post-Processor, \emph {Generator}}: We use an extension of the original UNet architecture~\cite{ronneberger2015u} as our generator and an extension of the perceptual critic used in~\cite{10222397}. Our UNet uses 5 downsampling($\div$2) and upsampling($\times2$) stages each. 

In our generator, we use a window of three RGB patches ($x_{t-1},x_t,x_{t+1}$) centered around time $t$. Patches $x_{t-1}$ and $x_{t+1}$ are motion compensated with respect to $x_t$ using DeepFlow\cite{6751282}.Prior to feeding the input sequence of frames into the UNet architecture (Generator), we concatenate the frames along a new dimension and perform a 3D convolution. After, the features are transformed into 3D tensor representations and then processed by the UNet using 2D convolutions. A feature map denoting the degredation level is concatenated to the input to the UNet. In our experiments, we used LPIPS to denote the level of degredation.

An encoder-decoder architecture was used to extract features from the nearest intra-coded frame. While the nearest intra-coded frame may not directly represent the current frame, it is expected to provide valuable contextual information. The inputs to each feature block from the I-frame encoder-decoder is concatenated with $x_i^{\delta t}$ using the time interval ($\Delta$ measured in frames) between the current frame and the nearest I-frame. We employ $x_i^{\Delta t} = e^{-0.02(|\Delta|)}$ hence features extracted from an I-frame near to the current frame are given a higher degree of relevancy. Note $0 \leq x_i^{\Delta t} \leq 1$. 

\textbf{Neural Post-Processor, \emph {Critic}}: Our architecture for our critic network is directly based on our previous work~\cite{10222397}. This critic network splices internal features from EfficientNetB3 as a preliminary input. The final output of the critic is a tensor that is 256 elements long which represents the learnt quality of the input patch. 

\section{Experiment}

\textbf{Data Set}: The data set used in the experiment consisted of a collection of 292 videos with varying resolutions. A set of 30 videos were set as validation and the remaining 262 videos were used as training. Compressed versions of these videos were created at the original resolution and multiple downsampled resolutions ($\div2$, $\div4$, $\div8$) using the encoding procedure in Algorithm~\ref{alg:bitrate} for target bitrates $50$kb/s and $500$kb/s. All compressed videos were then brought back to the original resolution and 100 patches of size $128\times128$ were then randomly sampled per video. 

\textbf{Loss Functions}: We train the generator and critic networks individually until they converge. Subsequently, both networks are linked together, and joint training is performed in an adversarial manner.

To train the generator to an initial stable state, we employ a loss of Mean Squared Error (MSE) for 15 epochs and then a loss function of $\Lagr_{gen}^{s}$ comprising of MSE and Learned Perceptual Image Patch Similarity (LPIPS) between the restored patch $\hat{x}_t$ and reference patch $y_t$ for a further 10 epochs: 
 $\Lagr_{gen}^{s} = MSE(\hat{x}_t, y_t) + LPIPS(\hat{x}_t, y_t) $. The inclusion of LPIPS in the loss function has shown to have improved perceptual results in imaging tasks~\cite{Jo_2020_CVPR_Workshops, Liu_2021_CVPR}.

Following individual training of the generator, we proceed with joint training using a combination of MSE, LPIPS, a feature loss $\Lambda(\cdot,\cdot)$ and the RaGAN-GP adversarial loss. The feature loss is based on the difference of the output of EfficientNetB3 layer 3 between restored and reference representations. The use of this loss function in the adversarial stage has been shown to give perceptually pleasing results~\cite{wang2021real, 9460113}. It encourages the generator to capture more fine-grained details and high-level image characteristics.  

The adversarial losses used for the generator and critic network are as follows, where GP represents the gradient penalty and is calculated as shown in~\cite{jolicoeur2018relativistic} and $\lambda_1$, $\lambda_2$ are weighting terms.
\begin{align}
\mathcal{L}_{crit} &= l_{b}(C(\hat{x}_t) - \overline{C(y_t)}, 0) + l_{b}(C(y_t) - \overline{C(\hat{x}_t)}, 1) + GP   \\
\mathcal{L}_{gen} &=  \lambda_1[l_{b}(C(\hat{x}_t) - \overline{C(y_t)}, 1) + l_{b}(C(y_t) - \overline{C(\hat{x}_t)}, 0)] \nonumber \\   &\phantom{=}+ MSE(\hat{x}_t, y_t) + \lambda_2LPIPS(\hat{x}_t, y_t) + 10\Lambda(\hat{x}_t, y_t)
\end{align}

Here $C(\cdot)$ is the output of the critic, $l_b(\cdot, \cdot)$ is the binary cross-entropy loss. and $\overline{(\cdot)}$ denotes the mean. $\lambda_2$ is set to 1000 for training, and $\lambda_1$ is set to 75, 125, 200, and 225 for training models that have no downsampling, 2x, 4x and 8x downsampled input data respectively. 

\textbf{Experimental Context}: We train eight post-processing models that each use data that are either not downsampled, downsampled by a factor of 2, 4, 8 respectively across two bitrates [50kb/s, 500kb/s]. We compare against the baseline libaom-av1 encoder and report improvements among multiple traditional and deep feature based metrics.

The Adan optimizer was employed with a learning rate of 1e-06 for individual training of the generator. After the individual training phase, both the generator and critic networks were trained together using the RaGAN-GP algorithm. The hyperparameter $n_{critic}$ and  was set to 2. The $n_{critic}$ parameter controls the ratio of training steps the critic undertakes relative to the generator to balance the training dynamics. Training was done on an NVidia GeForce RTX 3090 (24GB VRAM) using the Tensorflow library. 

\begin{table}[]
\centering
\caption{\em Averages of different metrics across the validation set. Post-processing results in a notable increase in performance for VMAF, LPIPS, DISTS and KID while having little negative impact on PSNR. Improvements across DISTS and VMAF are notable, while PSNR remains very close to the diagonal line.}
\label{tab:results_means}
\begin{adjustbox}{width=1\textwidth}
\small
\begin{tabular}{@{}lllllllllll@{}}
\toprule
                         & $PSNR_Y\uparrow$ & $PSNR_{CBCR}\uparrow$ & $PSNR_Y^{HVS}\uparrow$ & $MSSIM\uparrow$ & $CIEDE2000\downarrow$ & $CAMBI\downarrow$ & $VMAF\uparrow$  & $LPIPS\downarrow$ & $DISTS\downarrow$ & $KID\downarrow$      \\ \midrule
AV1 (50kb/s)             & \underline{28.43} & \underline{39.72}    & \underline{24.20}    & \underline{0.846} & 32.47     & \underline{2.23}  & 23.22 & 0.422 & 0.253 & 1.43e-05 \\
Post-Processed (50kb/s)  & 28.34 & 39.26    & 24.12    & 0.844 & \underline{32.22}     & 2.47  & \underline{29.94} & \underline{0.401} & \underline{0.194} & \underline{1.23e-05} \\ \midrule
AV1 (500kb/s)            & \underline{35.77} & \underline{45.02}    & \underline{33.29}    & 0.963 & 39.04     & 1.56  & 71.06 & 0.229 & 0.094 & 2.25e-04 \\
Post-Processed (500kb/s) & 35.59 & 44.74    & 33.17    & \underline{0.964} & \underline{38.74}     & \underline{0.36}  & \underline{72.87} & \underline{0.223} & \underline{0.074} & \underline{1.87e-04} \\ \bottomrule
\end{tabular}
\end{adjustbox}
\end{table}

\section{Results}

Table~\ref{tab:results_means} shows the mean score of different metrics from the validation set when they are encoded with libaom-av1 and then enhanced with our neural post-processor. Our approach has lower $PSNR_Y$, $PSNR_{CBCR}$ and $PSNR_Y^{HVS}$ for both 50kb/s (-0.09dB, -0.46dB, -0.06dB) and 500kb/s (-0.18dB, -0.72dB, -0.12dB) scores for both target bitrates. While the deviation in PSNR is very subtle, the increase in other metrics such as VMAF, LPIPS, DISTS and KID is notable. This is further reinforced by a paired t-test done one VMAF and DISTS which shows the improvements are significant at a 0.01 level of significance.

\begin{figure}
    \centering
    \includegraphics[width=\textwidth]{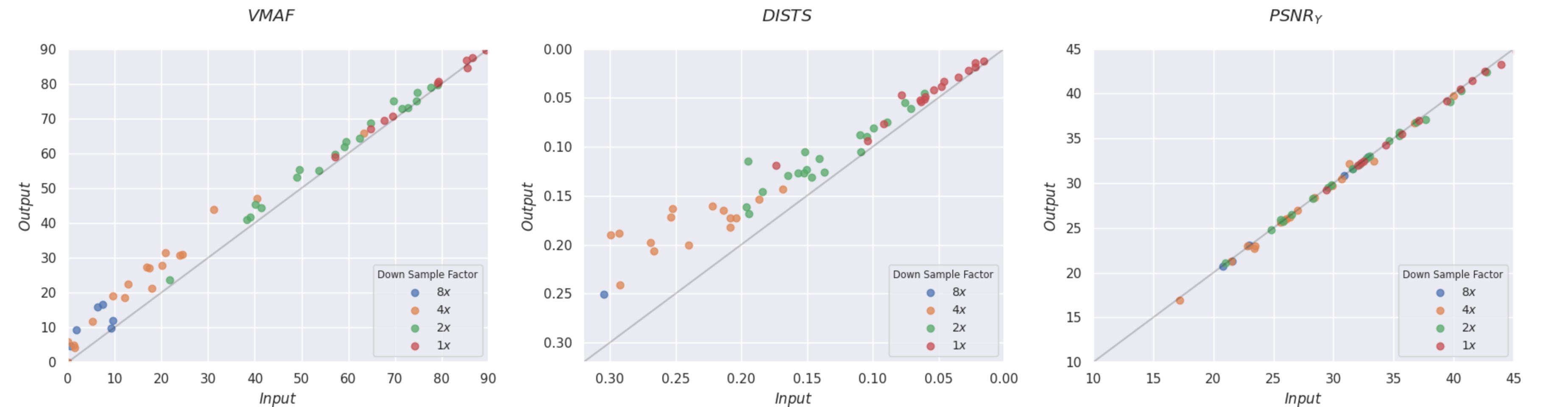}
    \caption{\em Metrics of videos before and after post-processing for both 50kb/s and 500kb/s. The x-axis shows the metrics at the input of the post-processor, and the y-axis shows the metrics of the content at the output of the post-processor. Positive vertical displacement (points above the solid diagonal line) indicates improvement.}
    \label{fig:input_output}
\end{figure}

Figure~\ref{fig:input_output} shows the input metric when compared to the output metric for videos from the validation set. It is notable that the our neural post-processor has higher improvements if the quality of the video is severly degraded (high DISTS or low VMAF). In particular, VMAF and DISTS shows large improvements (large +ve vertical displacement) in these areas, while the deviation around PSNR is minimal. This is due to our optimization criteria being deep feature focused over MSE/PSNR. 

\begin{figure}
    \centering
    \includegraphics[width=0.55\textwidth]{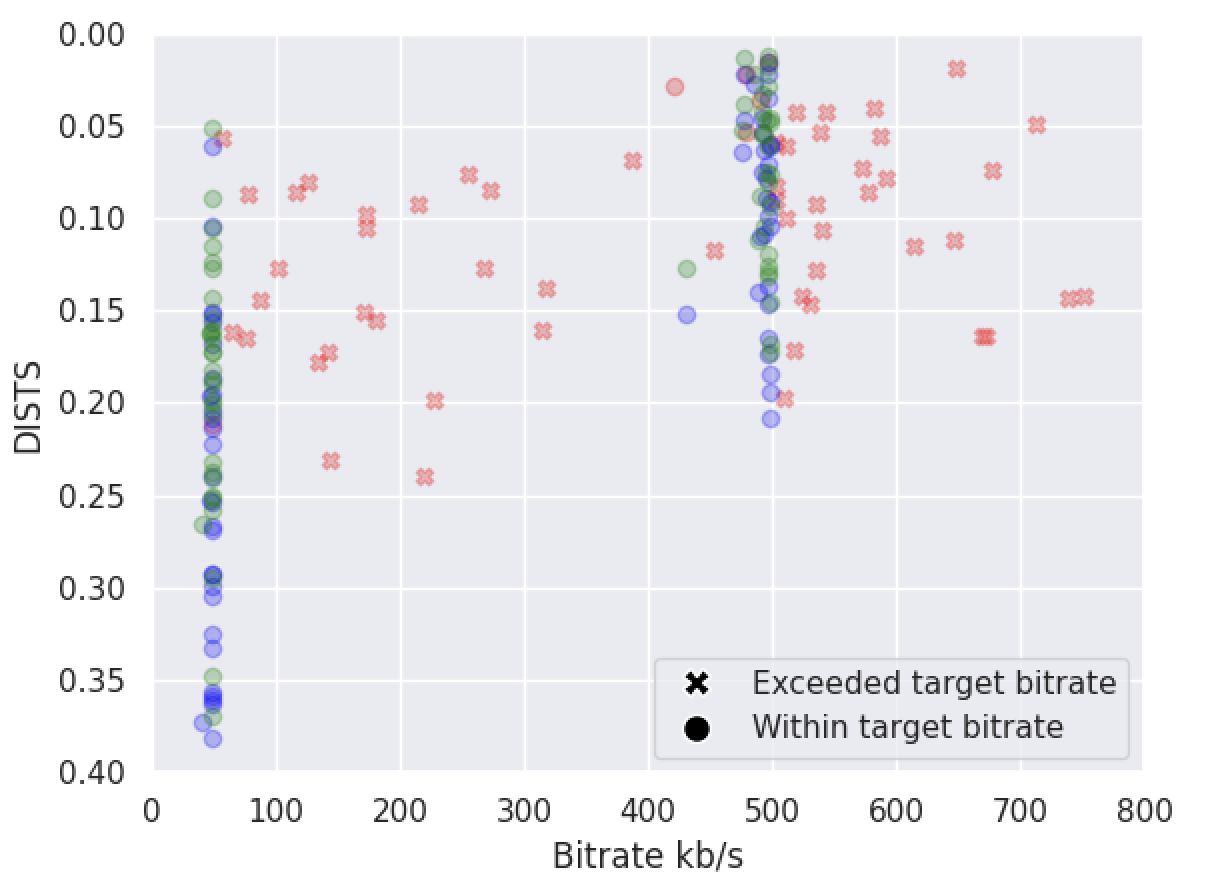}
    \caption{\em Scatterplot of DISTS compared to output bitrate for the validation set encoded with libaom-av1 (red points), our compression search scheme (blue points) and post-processing (green points). The majority of the red points have exceeded the target bitrate criteria [50kb/s, 500kb/s]. Using our compression scheme with our neural post-processor increases the quality of these videos are hight and the bitrate is within a given target.}
    \label{fig:scatterplot}
\end{figure}

Figure~\ref{fig:scatterplot} shows the difference between using libaom-av1 with a CBR setting of the target bitrate, using our improved compression pipeline and using our neural post-processor coupled with our compression pipeline. The majority of points ($90\%$) encoded with libaom-av1 exceeds the target bitrate. The mean, $\mu$, and standard deviation, $\sigma$, of the encoded versions using this method (red points) is [$\mu$: 245kb/s, $\sigma$: 184.1kb/s] and [$\mu$: 555.9kb/s, $\sigma$: 75.19kb/s]for a target of 50kb/s and 500kb/s respectively. Using our compression pipelines (blue points) results in the mean and standard deviation of the encoded versions to be [$\mu$: 48.33kb/s, $\sigma$: 1.74kb/s] and [$\mu$: 491.13kb/s, $\sigma$: 13.1kb/s]for a target of 50kb/s and 500kb/s respectively.

\begin{figure}
    \centering
    \includegraphics[width=0.91\textwidth]{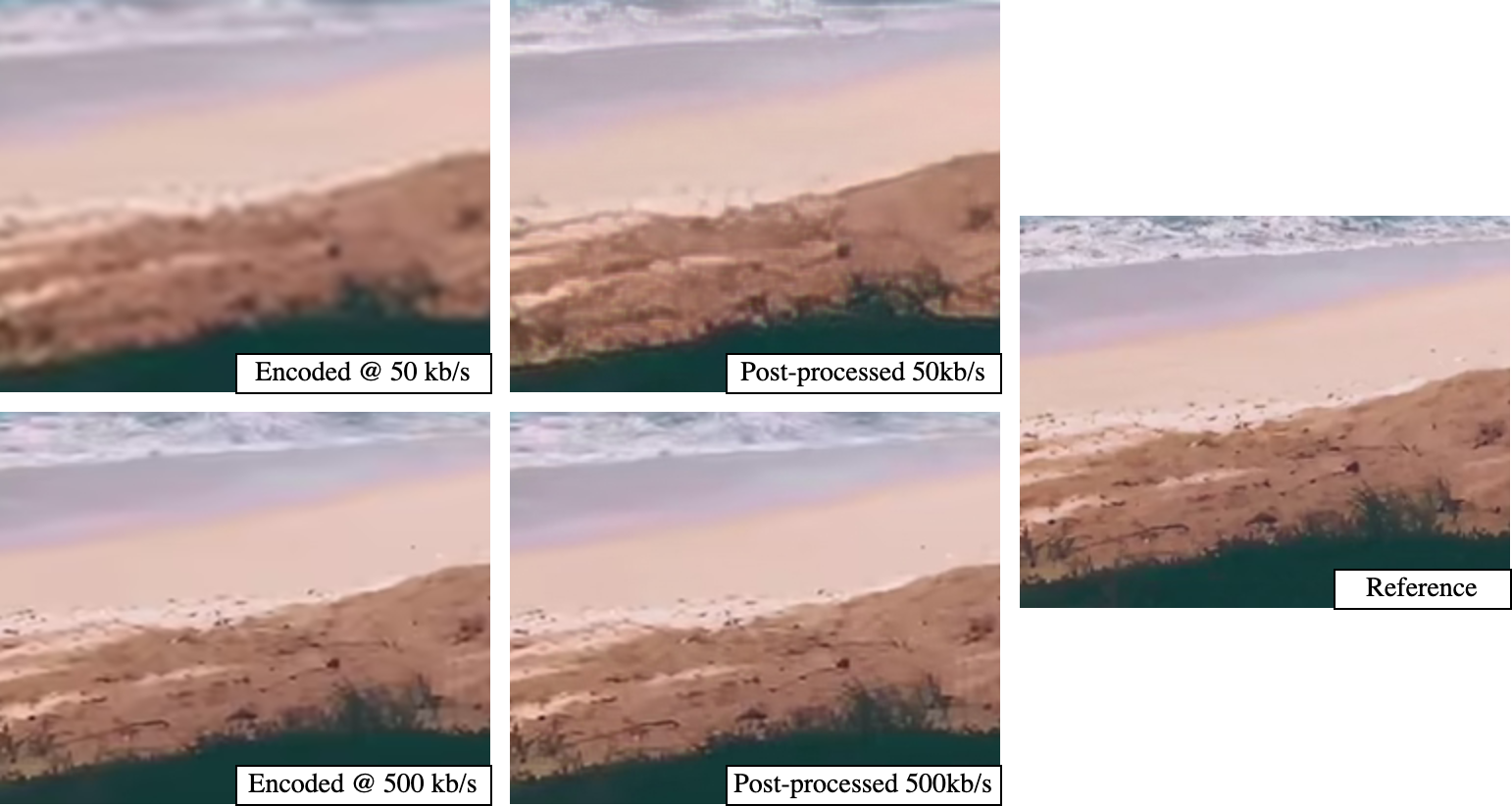}
    \caption{\em Reference video encoded @ 50kb/s, 500kb/s and then post-processed. Patches are extracted from videos with DISTS$\vert$VMAF of [0.22$\vert$17.365, 0.16$\vert$27.09, 0.06$\vert$69.47, 0.05$\vert$70.63] for encoded @ 50kb/s, post-processed @ 50kb/s, encoded @ 500kb/s and post-processed @ 500kb/s respectively.}
    \label{fig:demo}
\end{figure}

Figure~\ref{fig:demo} shows the visual difference between encoded and post-processed versions. There is much more detail with the post-processed 50kb/s version compared to the relevant encoded version. The difference between the post-processed 500kb/s version and the encoded 500kb/s is minimal, but that is due to the encoded version being very close to the reference patch as is. 

\textbf{Conclusion}: We present a method for encoding a video to a target bitrate which is then decoded and enhanced. The encoding algorithm uses the libaom-av1 encoder to generate candidate representations at different spatial resolutions. If a potential candidate is not found, then the temporal resolution is changed. Our post-processing enhancement network has been trained on four different input data types, that reflect the level of down-sampled required to achieve a given bitrate. Our results show that while we have lower PSNR than the reference encoder, we produce patches that are consistently better in perceptual metrics such as VMAF, LPIPS, DISTS and KID. 


\Section{References}
\bibliographystyle{IEEEbib}
\bibliography{refs}

\end{document}